\documentclass[%
 reprint,
 amsmath,amssymb,
 aps,
]{revtex4-2}

\usepackage{graphicx}
\usepackage{dcolumn}
\usepackage{bm}
\usepackage{hyperref}

\begin{document}

\setlength{\textfloatsep}{10pt}
\setlength{\abovecaptionskip}{10pt} 
\setlength{\belowcaptionskip}{10pt} 

\preprint{APS/123-QED}

\title{Josephson Traveling-Wave Parametric Amplifier with Inverse Kerr Phase Matching}

\author{M.T. Bell}
 \email{Matthew.Bell@umb.edu}
\affiliation{Engineering Department, University of Massachusetts Boston, Boston, Massachusetts, USA}
\date{\today}

\begin{abstract}
Superconducting traveling-wave parametric amplifiers (TWPA) have emerged as highly versatile devices, offering broadband amplification with quantum-limited noise performance. They hold significant potential for addressing the readout bottleneck in prototype quantum computers, enabling scalability. Key challenges with this technology include achieving sufficient gain with minimal gain ripple while maintaining low noise performance. Efficient phase matching between a weak signal and a strong pump over the entire length of the TWPA is critical to overcoming these challenges. We present an experimental demonstration of the inverse Kerr phase matching technique in a TWPA, first proposed in Ref. Phys. Rev. Appl. 4, 024014. This method addresses several limitations of conventional dispersion engineering approaches of phase matching in the four-wave mixing parametric process in TWPAs. Most notably the existence of an unusable region of gain near the pump frequency which typically corresponds to the region of most optimal phase matching and maximum gain. The inverse Kerr phase matching approach, allows for greater frequency separation between the region of optimal gain and pump, \textit{in situ} tunability of the pump, minimal gain ripple, and a compact footprint which reduces losses. A TWPA employing the inverse Kerr phase matching technique experimentally demonstrated 20 dB of gain over a 3 GHz instantaneous bandwidth, with a tunable bandwidth of 8 GHz, minimal gain ripple, and near quantum-limited noise performance, with 1.5 photons of added noise.
\end{abstract}

\maketitle


\section{\label{intro}Introduction}
Superconducting Traveling-Wave Parametric Amplifiers (TWPA) have recently received significant attention due to their remarkable versatility \cite{aumentado_review, Roch_twparev,twpascience,siddiqitwpa1,siddiqitwpa2,whitetwpa,delsing3wm,olegtwpa}. A key application of TWPAs is amplifying weak signals from superconducting quantum bits (qubits) at millikelvin temperatures, improving readout fidelities, and enabling multiplexed readout strategies crucial for quantum error correction in large-scale quantum systems \cite{oliver_guide,qubit1,wallraff100qubit,wallraffmulti}. Beyond this, TWPAs have been used in a variety of applications including, spin-based qubit readout \cite{spin_qubit1}, broadband microwave squeezing \cite{Rochsqueezing1,oliversqueezing1,Blaissqueeze1,revkerr_squeezing}, dark matter searches \cite{darkmatter1,darkmatter2,darkmatter3}, and readout of microwave single-photon detectors \cite{twpamkid1,twpamkid2}.

TWPAs are typically the first amplifier in the signal readout chain and plays a critical role in setting the overall system noise \cite{pozar}. TWPAs fabricated with low-loss materials have demonstrated near quantum-limited noise performance, adding approximately one photon of noise \cite{whitetwpa}. To improve system noise, TWPAs are designed with ~20 dB of gain to amplify weak signals above the noise floor of subsequent amplifiers in the readout chain. TWPAs have demonstrated wide bandwidth amplification of several GHz and high saturation powers which allows for the multiplexed readout of quantum circuits. TWPA's are receiving considerable interest as a source of non-classical states of light \cite{Blaissqueeze1,oliversqueezing1,Rochsqueezing1,revkerr_squeezing} and as a non-reciprocal amplifier with sufficient reverse isolation \cite{aumentadotwpaiso,Rochtwpaiso,twpaiso_counterpropagating,amp_nonrecip2}, eliminating the need for bulky ferrite-based isolators and promoting scalability of readout.

TWPAs amplify weak microwave signals via a parametric wave-mixing process, where a strong pump tone $\omega_p$ interacts with a weak signal $\omega_s$, generating an idler tone $\omega_i$ in a nonlinear medium. There are two wave mixing parametric processes utilized for amplification: three wave mixing (3WM) and four wave mixing (4WM). These parametric processes facilitate energy exchange between the coupled modes resulting in amplification.  For efficient amplification, both energy (momentum) conservation $\omega_p=\omega_s+\omega_i$ ($k_p=k_s+k_i$)  and $2\omega_p=\omega_s+\omega_i$ ($2k_p=k_s+k_i$) for 3WM and 4WM parametric processes respectively must be satisfied for propagating waves with wave number $k_n$, where $n=\{s,i,p\}$ in the nonlinear medium \cite{agrawal2013nonlinear,boyd2008nonlinear}. This nonlinear medium can consist of transmission lines made from Josephson junctions \cite{twpascience,lefthandedtwpa}, superconducting quantum interference devices (SQUIDs) \cite{Rochphotonic}, superconducting nonlinear asymmetric inductive elements (SNAIL) \cite{Rochrevkerr}, or high-kinetic inductance disordered superconductors \cite{3wm_kin,Eomtwpa}.

Several phase-matching solutions have been proposed for both 3WM and 4WM parametric processes, but these often involve complex dispersion engineering strategies. In 3WM, dispersion engineering circuits are introduced to suppress unwanted second harmonic generation, which can dominate the parametric process \cite{3wm_kin,Eomtwpa}. In 4WM, dispersion engineering is required to maintain phase matching between the signal, idler, and pump as they propagate through a nonlinear transmission line. These waves are affected by nonlinearities and chromatic dispersion in the transmission line resulting in phase mismatch. Current approaches to correcting phase mismatch include the introduction of photonic bandgaps \cite{Rochphotonic,Eomtwpa} or the use of resonant circuits \cite{oliversqueezing1,twpascience,whitetwpa,siddiqitwpa2} to introduce a dispersive feature into the transmission of the TWPA for which to tune $\omega_p$ around. However, these methods have limitations, including restricted tunability of $\omega_p$, the existence of a unusable region of the bandwidth around the pump, and gain ripples due to impedance mismatches introduced by a dispersive feature \cite{Rochrevkerr}. These approaches have tradeoffs which ultimately limit the versatility of the TWPA.

In this article, we demonstrate a Josephson Junction Traveling-Wave Parametric Applifier (JTWPA) which utilizes a novel 4WM phase-matching approach, first introduced in Ref. \cite{belltwpa,bellpatent}. This approach leverages the balance between the phase mismatch introduced by an inverse sign third-order Kerr nonlinearity and the chromatic dispersion of the JTWPA transmission line. This method enables efficient 4WM parametric amplification without the need for traditional dispersion engineering techniques. By eliminating the complexity of such approaches, we avoid common drawbacks with conventional dispersion engineering approaches: gaps in the gain bandwidth, gain ripples, and limited tunability in $\omega_p$. The inverse Kerr phase-matching technique produces a gain bandwidth profile where the phase-matched maximum gain is positioned far from $\omega_p$. This separation allows for better isolation of the pump from the signal, in contrast to conventional dispersion engineering techniques where phase matching and maximum gain occurs at $\omega_p$ \cite{siddiqitwpa2,twpascience}.

\section{\label{twpadesign}JTWPA Design}
\begin{figure}
\includegraphics[scale=0.94]{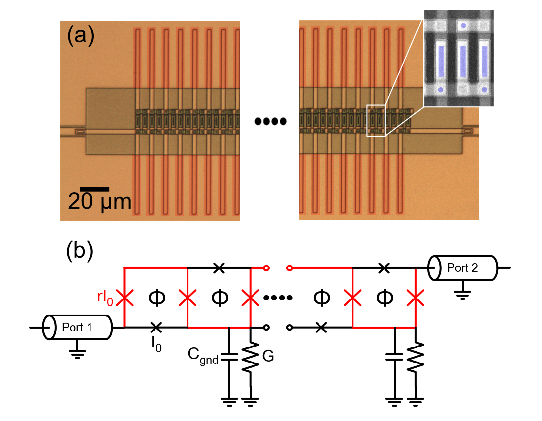}
\caption{\label{fig1}(a) An optical microscope image showing the beginning and end of the JTWPA, a coplanar transmission line composed of coupled asymmetric SQUIDs. The red squares represent the parallel plate capacitors that form $C_{\rm gnd}$. The inset shows a scanning electron microscope image of a unit-cell, with the large and small Josephson junctions indicated by blue squares and circles, respectively. (b) The circuit diagram of the JTWPA transmission line, terminated with 50 $\Omega$ impedance at ports 1 and 2. The meandered backbone of large Josephson junctions in the transmission line is highlighted in red. The asymmetric SQUIDs are tuned via an applied magnetic flux $\Phi$. Conductance $G$ accounts for the dielectric loss in the transmission line.}
\end{figure}

The JTWPA investigated consists of a co-planar transmission line of coupled asymmetric SQUIDs \cite{bellkerr,bellsuperind,belltwpa}. An alternative approach can be realized with SNAIL elements \cite{Rochrevkerr}. The JTWPA structures were fabricated at the SeeQC superconducting electronics foundry \cite{seeqc} and are shown in the optical microscope image in Figure \ref{fig1}(a). The electrical circuit diagram of a unit-cell of the JTWPA is presented in Figure \ref{fig1}(b), where each unit-cell has a length $a$ = 10 $\mu$m. Each unit-cell consists of two coupled asymmetric SQUIDs: one arm contains a single small Josephson junction (denoted by a black "×") with critical current $I_0$, while the other arm contains two larger Josephson junctions (denoted by red "×") with critical current $rI_0$, where $r$ represents the ratio in the areas between the large and small junctions. The small and large junctions have shunting capacitances of $C_0$ and $rC_0$, respectively. The asymmetric SQUIDs are coupled through the large junctions, forming a meandered backbone, as indicated in red. The red squares in Figure \ref{fig1}(a) highlight the parallel plate capacitors to ground $C_{\rm gnd}$, which are essential for maintaining a 50 $\Omega$ impedance in the transmission line. To minimize insertion loss, the dielectric material of these grounding capacitors is low-loss SiNx.  Each asymmetric SQUID is threaded by a magnetic flux $\Phi$, which allows for tuning of the JTWPA's operating point. Two nominally identical JTWPAs with different transmission line lengths $l = 350a$ (3.5 mm) and $l = 865a$ (8.65 mm) are measured.

The coupled SQUID design enables both a flux-tunability of the linear and third-order nonlinear component in the expansion of the current-superconducting phase $\varphi$ relation of the JTWPA unit-cells:
\begin{eqnarray}\label{eq1}
I(\varphi)=I_0\left[\frac{r}{2}+2\cos\left(2\pi\frac{\Phi}{\Phi_0}\right)\right]\varphi-\nonumber\\
\frac{I_0}{3}\left[\frac{r}{16}+\cos\left(2\pi\frac{\Phi}{\Phi_0}\right)\right]\varphi^3,
\end{eqnarray}

\noindent where $\Phi_0 = h/(2e)$ represents the magnetic flux quantum. The flux tunability of the unit-cell’s linear component allows for \textit{in situ } tuning of the characteristic impedance of the JTWPA. A distinctive feature of this design is its ability to tune the third-order nonlinearity over a wide range via $\Phi$, and even invert its sign \cite{bellkerr}. This capability enables inverse Kerr shift phase matching for a 4WM process in the JTWPA while maintaining a zero second-order nonlinearity, which suppresses unwanted 3WM parametric processes.

The propagation of electromagnetic waves along the $z$ direction in the JTWPA, with wavelengths $\lambda \gg a$, is governed by a wave equation \cite{belltwpa}:

\begin{eqnarray}\label{eq2}
\frac{1}{L}\frac{\partial^2\varphi}{\partial z^2}+C_0\left(\frac{r}{2}+2\right)\frac{\partial^4\varphi}{\partial t^2 \partial z^2}-C_{\rm gnd}\frac{\partial^2\varphi}{\partial t^2}-\nonumber\\
G\frac{\partial \varphi}{\partial t}-\gamma\frac{\partial}{\partial z}\left[\left(\frac{\partial \varphi}{\partial z}\right)^3\right]=0,
\end{eqnarray}

\noindent where $L=L_0/[r/2+2\cos(2\pi\Phi/\Phi_0)]$, $L_0=\Phi_0/(2\pi I_0)$, and $\gamma=1/(3\varphi_0^2L_0)[r/16+\cos(2\pi\Phi/\Phi_0)]$ is the Kerr nonlinearity. Dielectric loss in the JTWPA is modeled by a frequency dependent per unit-cell conductance to ground $G=\omega C_{gnd} \tan\delta$ where $\tan\delta$ is the loss tangent of the SiNx dielectric. The chromatic dispersion relation is obtained in the absence of nonlinearities and loss to be 

\begin{eqnarray}\label{eq3}
k(\omega, \Phi)=\frac{\omega\sqrt{L_0 C_{\rm gnd}}}{\sqrt{\left[\frac{r}{2}+2\cos\left(2\pi\frac{\Phi}{\Phi_0}\right)\right]-\omega^2L_0C_0\left(\frac{r}{2}+2\right)}}.
\end{eqnarray}

\noindent A solution to Eq. \ref{eq2} involves four traveling waves. In the degenerate case, it is two pump waves with equal $\omega_p$, a signal $\omega_s$, and a generated idler $\omega_i=2\omega_p-\omega_s$. A set of coupled mode equations can be derived \cite{belltwpa, siddiqitwpa1} to describe the interaction of the signal and idler waves under the influence of a stiff pump, 

\begin{eqnarray}\label{eq4}
\frac{\partial a_s}{\partial z}-i\kappa_s a_i^*e^{i\kappa z}=0,\nonumber\\
\frac{\partial a_i}{\partial z}-i\kappa_i a_s^*e^{i\kappa z}=0,
\end{eqnarray}

\noindent where $a_n$ is a complex amplitude of the mode $n = \{s,i,p\}$. The couplings between the signal and idler in Eq. \ref{eq4} are

\begin{eqnarray}\label{eq5}
\kappa_s=\frac{3\gamma k_p^2k_s k_i(2k_p-k_i)|A_p|^2}{8\omega_s^2C_{\rm gnd}},\nonumber\\
\kappa_i=\frac{3\gamma k_p^2k_s k_i(2k_p-k_s)|A_p|^2}{8\omega_i^2C_{\rm gnd}},
\end{eqnarray}

\noindent where $A_p$ is the pump amplitude. A standard solution for Eq. \ref{eq4} is the power gain \cite{boyd2008nonlinear}:

\begin{eqnarray}\label{eq6}
G_s=\text{cosh}^2(gz)+\frac{\kappa^2}{4g^2}\text{sinh}^2(gz),
\end{eqnarray}

\noindent where the exponential gain factor is

\begin{eqnarray}\label{eq7}
g=\sqrt{\kappa_s\kappa_i-\left(\frac{\kappa}{2}\right)^2}.
\end{eqnarray}

\noindent The total phase mismatch $\kappa = \Delta k+\alpha_{nl}$ is a combination of phase mismatch due to chromatic dispersion $\Delta k=k_s+k_i-2k_p$ and self-phase (SPM) and cross-phase (XPM) modulation $\alpha_{nl}=\alpha_s+\alpha_i-2\alpha_p$ where

\begin{eqnarray}\label{eq8}
\alpha_{s,i}=\frac{3\gamma k_{s,i}^2k_p^2|A_p|^2}{4C_{\rm gnd}\omega_{s,i}^2},\nonumber\\
\alpha_{p}=\frac{3\gamma k_p^5|A_p|^2}{8C_{\rm gnd}\omega_{p}^2}.
\end{eqnarray}

\noindent Critical to this approach of phase matching is the ability to tune the sign and magnitude of $\alpha_n$ through the flux tunable nonlinearity $\gamma$ and $A_p$ respectively, to compensate $\Delta k$ and achieve phase matching with $\kappa=0$. When phase matched, $g\approx|\alpha_p|$ and the JTWPA power gain Eq. \ref{eq6} is exponential in $z$ and $|A_p|^2$ as $G_s\approx\exp(2|\alpha_p| z)/4$. There exists a region of phase mismatch near the pump where $\Delta k\approx 0$ and $\kappa\approx2\alpha_p$, in this region the power gain is limited to a quadratic dependence $G_s\approx 1+(\alpha_p z)^2$ on $z$ and $|A_p|^2$.

\section{\label{results}Results}

In this work two nominally identical JTWPAs differing primarily in length with circuit design parameters in Table \ref{table1}, were experimentally characterized in a dilution refrigerator at 20 mK with the measurement setup shown in Figure \ref{msetup}. An external superconducting solenoid was used to apply $\Phi$ to the JTWPA. The critical current $I_0$ was estimated using the Ambegaokar-Baratoff relation \cite{tinkham}, based on room temperature normal state resistance measurements from test junctions fabricated in the same process run. The permittivity of the SiNx dielectric was characterized using test resonators to determine $C_{\rm gnd}$. The Josephson junction capacitance $C_0$ was calculated using 40 fF/$\mu m^2$. The $r$ value was defined in the artwork submitted to the foundry \cite{seeqc}.
\begin{table}
\caption{\label{table1}%
A table of JTWPA parameters. Numbers outside parenthesis are parameters by design. Numbers inside parenthesis are obtained from fitting to experimental data.
}
\begin{ruledtabular}
\begin{tabular}{cccccc}
\textrm{JTWPA}&
\textrm{$l$ ($a$)}&
\textrm{r}&
\textrm{$I_0 (\mu A)$}&
\textrm{$C_0 (fF)$}&
\textrm{$C_{gnd} (fF)$}\\
\colrule
A & 865 & 6(6.2) & 1.2(1.25) & 40(45) & 110(115)\\
B & 350 & 6(6.2) & 1.2(1.25) & 40(45) & 110(115)\\
\end{tabular}
\end{ruledtabular}
\end{table}

\begin{figure}
\includegraphics[scale=0.55]{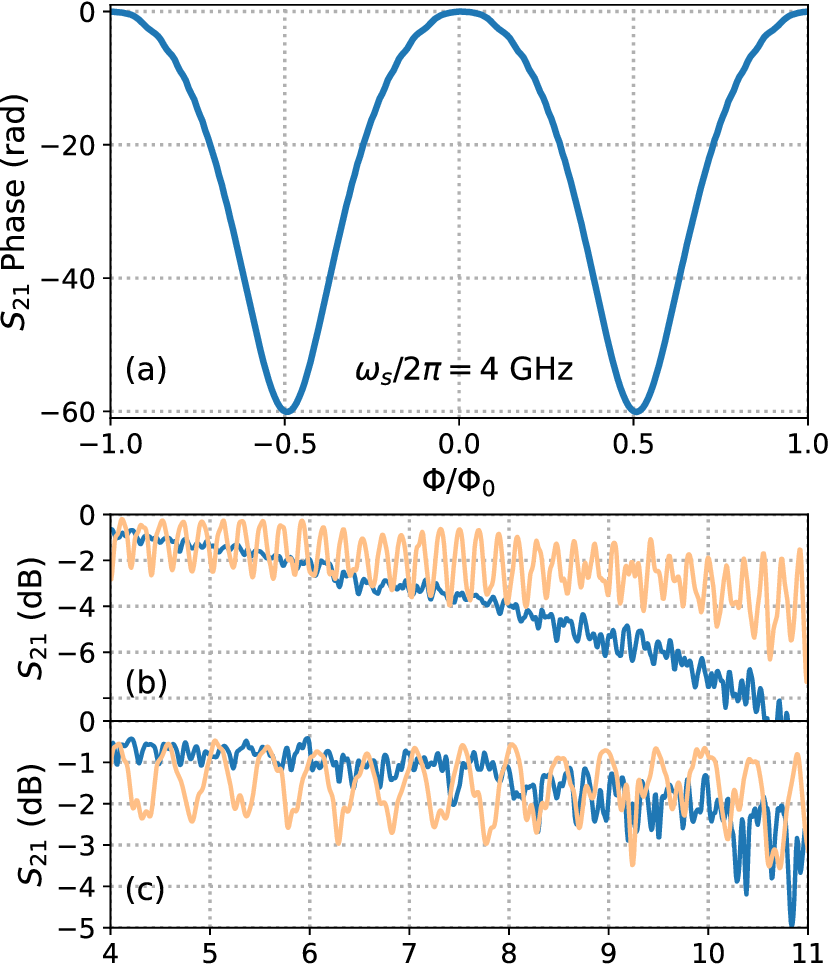}
\caption{\label{fig2}(a) Transmission $\angle S_{21}$ versus $\Phi/\Phi_0$ at $\omega_s/2\pi = 4$ GHz for JTWPA B. (b) and (c) Transmission measurements of JTWPA A and B respectively for $\Phi/\Phi_0 = 0$ (orange lines) and $\Phi/\Phi_0 = 0.5$ (blue lines).}
\end{figure}

The JTWPAs were first characterized with transmission measurements in the low-power linear regime, with no pump applied, to determine the circuit's electrical parameters. Transmission measurements were calibrated against an on-chip thru-line, housed in an identical sample box, which could be switched in and out of the readout chain using cryogenic RF switches. The thru-line calibration measurements provide a transmission magnitude and phase reference for dispersion measurements. Transmission measurements of the phase (in radians) as a function of $\Phi$ at $\omega_s/2\pi = 4$ GHz are shown in Figure \ref{fig2}(a), illustrating the periodic dependence of the propagation constant on $\Phi$. Transmission measurements for JTWPAs A and B at two different tuning, $\Phi/\Phi_0=0$ (orange line) and 0.5 (blue line), are shown in Figures \ref{fig2}(b) and \ref{fig2}(c), respectively. Both JTWPAs were designed with a critical current $I_0$ and ground capacitance $C_{\rm gnd}$, resulting in a characteristic impedance of 50 $\Omega$ at $\Phi/\Phi_0=0.5$. The overall insertion loss at 6 GHz was measured to be approximately 2.1 dB for JTWPA A and 0.9 dB for JTWPA B, reflecting the low-loss properties of the SiNx dielectric used in $C_{\rm gnd}$. The measured insertion loss suggests a SiNx dielectric loss tangent $\tan\delta\sim 0.0027$ higher than expected when compared to the $\tan\delta\sim 0.0005$ obtained from test resonator measurements. The increased losses in the dielectric of the grounding capacitors may result from their higher aspect ratio, highlighted in red in Figure \ref{fig1}(a), compared to the lower aspect ratio parallel plate capacitor measured in the lumped element test resonator structures. Despite the higher-than-expected loss, the JTWPAs exhibit lower insertion losses compared to many reported in literature, which are often fabricated using higher-loss dielectrics \cite{twpascience,Rochrevkerr}.

Low-loss JTWPAs are more susceptible to standing wave ripple in transmission due to variations in distributed circuit parameters and minimal damping to suppress them \cite{paramvariations1,paramvariations2,paramvariations3}. For the JTWPAs studied, a clear standing wave in the transmission is observed in Figure \ref{fig2}(b),(c) (orange line) caused by an intentional impedance mismatch with a tuning of $\Phi/\Phi_0=0$. Under impedance-matching conditions at $\Phi/\Phi_0=0.5$, Figure \ref{fig2}(b),(c) (blue line), the standing wave ripple was minimal, indicating a highly uniform fabrication process with little spread in Josephson junction parameters. Measurements of over 100 test Josephson junctions normal state resistance revealed a critical current variation with a standard deviation below 2\% for junctions within a 10 mm distance on the wafer. In higher-loss JTWPAs, standing wave effects are often masked by dielectric losses.

\begin{figure}[b]
\includegraphics[scale=0.58]{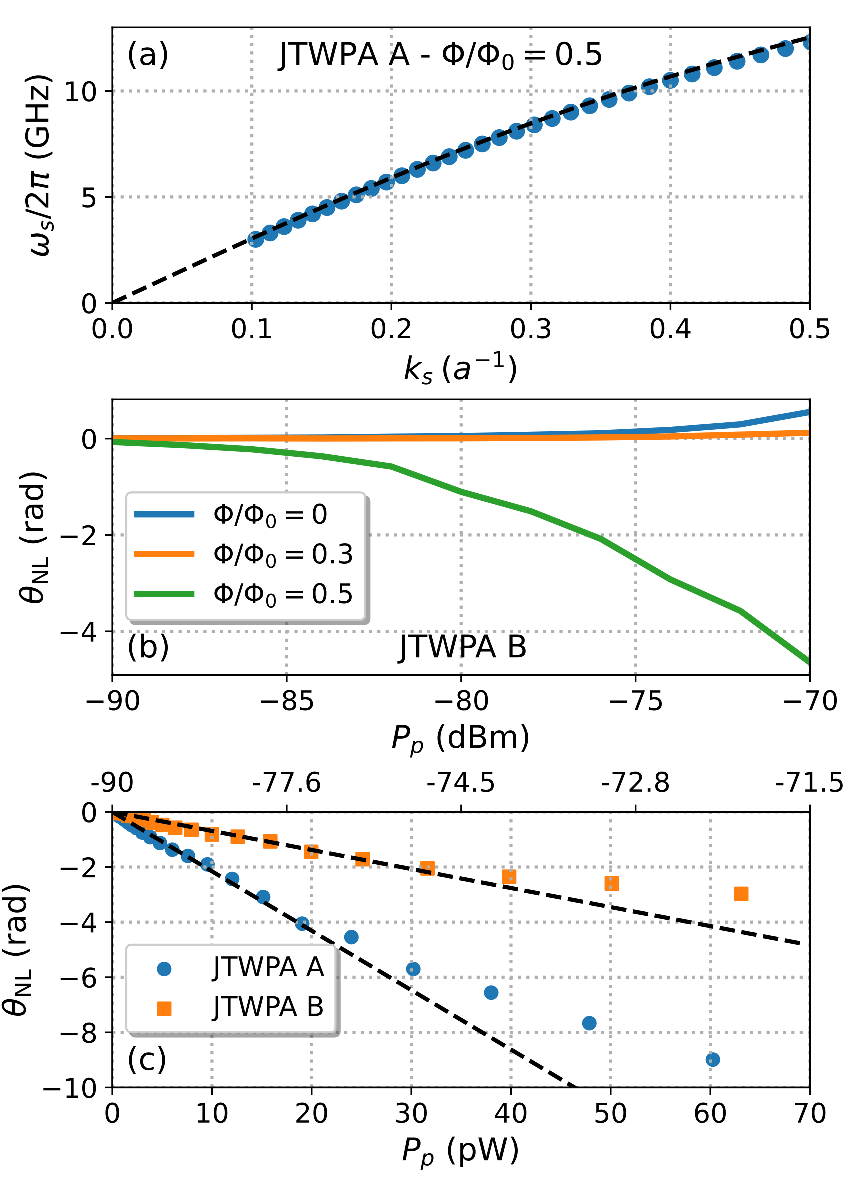}
\caption{\label{fig3}(a) Measured chromatic dispersion of JTWPA A, dashed line is a fit to Eq. \ref{eq3}. (b) Measured $\theta_{\rm NL}=\angle S_{21}(\omega_p/2\pi=6 \text{ GHz})$ versus $P_p$ for JTWPA B for $\Phi/\Phi_0$: 0 (blue line), 0.3 (orange line), and 0.5 (green line) shows how the JTWPA Kerr nonlinearity can be tuned over a wide range and invert sign. (c) Pump $\theta_{\rm NL}$ for both JTWPA A and B versus $P_p$ in linear scale (bottom axis) for comparison with Eq. \ref{eq8} and in dBm (top axis). Dashed lines are expected $\theta_{\rm NL}=\alpha_p l$ dependence on $P_p$. For low $P_p$, experimental data agrees with theory. For higher $P_p$ there is a deviation from linear dependence indicating a practical limit on $P_p$.}
\end{figure}

Measurements of the calibrated transmission phase $\angle S_{21} = k_sl$ versus $\omega_s$ at $\Phi/\Phi_0 = 0.5$, were used to extract the frequency versus wavenumber $k_s$ shown in Figure \ref{fig3}(a). The dashed line is a fit to Eq. \ref{eq3}, which is used to determine the circuit parameters listed in Table \ref{table1}. The transmission measurements in the linear operating region align well with design parameters. The observed chromatic dispersion in the phase response versus frequency is attributed to a plasma frequency at $\omega_0/2\pi \sim 30$ GHz, introduced by the Josephson junction capacitance. The chromatic dispersion leads to a non-zero $\Delta k$, which is necessary for the inverse Kerr phase matching approach.

A distinct feature of this JTWPA circuit is its Kerr nonlinearity, which can be tuned in both amplitude and sign with $\Phi$. Figure \ref{fig3}(b) illustrates this tunability by measuring the phase shift of the pump $\theta_{\rm NL} = \angle S_{21}(\omega_p/2\pi=6 \text{ GHz})$ versus pump power $P_p$ for various $\Phi/\Phi_0$. Theoretically $\theta_{\rm NL}=\alpha_pl$. For $\Phi/\Phi_0=0$ (blue line), the Kerr nonlinearity has the same sign and direction of phase shift as that found in the expansion of the sinusoidal Josephson junction current-phase relation. At $\Phi/\Phi_0=0.3$ (orange line), the JTWPA is in the Kerr-free region, and the phase response does not exhibit to cubic order dependence on $P_p$. Near $\Phi/\Phi_0=0.5$, the Kerr nonlinearity changes sign (direction of phase shift), and its amplitude is tunable with $\Phi/\Phi_0$ over a wide range. This sign inversion in the Kerr nonlinearity is crucial for phase matching, as it balances the non-zero $\Delta k$ with $\alpha_{nl}$ due to SPM and XPM between signal, idler, and pump.

To achieve phase matching, the amplitude of the Kerr nonlinearity with an inverse sign is tuned so that phase matching conditions are met for a signal, idler, and pump. The dependence of this amplitude is assessed through $\theta_{\rm NL}$ versus $P_p$ shown in Figure \ref{fig3}(c) for JTWPAs A (blue circles) $\Phi/\Phi_0 = 0.475$ and B (orange squares) $\Phi/\Phi_0 = 0.45$. For low $P_p$, $\theta_{\rm NL}$ is linear with $P_p$ in agreement with Eq. \ref{eq8}, shown as a dashed line. For $P_p$ exceeding 16 pW (-78 dBm) for JTWPA A and 32 pW (-75 dBm) for JTWPA B, $\theta_{\rm NL}$ deviates from linear behavior. This deviation indicates a limit for $P_p$, which ultimately limits the gain of the JTWPA.

The expected gain of JTWPAs is typically estimated by numerically solving a series of coupled mode equations (CME) that account for a signal, idler, and pump \cite{siddiqitwpa1}. Such numerical simulations generally overestimate the gain due to several factors, including the truncation of higher-order terms in the expansion of the sinusoidal current-phase relation of the Josephson junctions, unaccounted parametric processes, losses, impedance mismatches, and variations in distributed circuit parameters \cite{twpasim3}. To address these limitations, numerical time-domain analysis of the JTWPA’s voltage response using circuit simulators has become common \cite{twpasim1,twpasim2,twpasim3,twpasim4,twpasim4_pumpharmonics,twpasim5}. Circuit simulators can incorporate variations in JTWPA parameters, losses, and complex frequency-dependent impedance environments to more accurately predict the JTWPA gain. Circuit simulators do have limitations in that they cannot account for unknown effects in experiment which may deplete the pump and overall phase matching efficiency and overestimate the gain.

\begin{figure}
\includegraphics[scale=0.58]{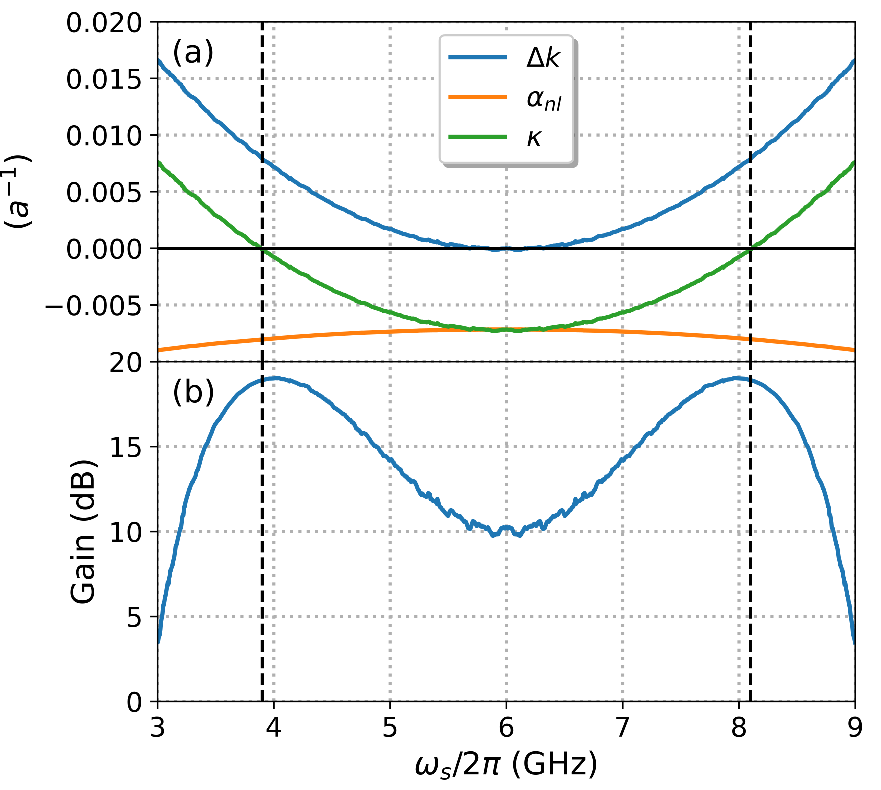}
\caption{\label{fig4}Calculation of the phase mismatch and the gain versus $\omega_s/2\pi$ from experimental transmission measurements of JTWPA A at $\omega_p/2\pi = 6$ GHz. (a) Measured chromatic dispersion $\Delta k$ (blue line), nonlinear phase mismatch due to SPM and XPM $\alpha_{nl}$ (orange line), and total mismatch $\kappa$ (green line). (b) Calculated gain from Eq. \ref{eq6}, utilizing only experimental data with no fitting parameters. Vertical dashed lines indicate phase matching condition where maximum gain occurs.}
\end{figure}

Here, the gain of JTWPA A is calculated using only measured propagation constants, both in the linear and nonlinear regions of operation when the pump is applied. From the linear response of the propagation constant, the chromatic dispersion $\Delta k$ versus $\omega_{s}$ for $\omega_p/2\pi = 6$ GHz is determined, as shown by the solid blue line in Figure \ref{fig4}(a). The effects of SPM and XPM, represented by $\alpha_{nl}$, are determined from nonlinear measurements of the propagation constant response to larger $P_p$, as shown by the solid orange line for $P_p$= -78±1.5 dBm. The uncertainty in $P_p$ is due to an incomplete understanding of line losses in our system at low temperatures. By combining the phase mismatch response from the chromatic dispersion and the SPM and XPM effects, the total phase mismatch $\kappa=\Delta k + \alpha_{nl}$ is obtained, as shown by the solid green line in Figure \ref{fig4}(a). By utilizing the experimentally measured $\Delta k$, $\alpha_{nl}$, and $\kappa$ and Eqs. \ref{eq5}, \ref{eq6}, \ref{eq7}, and \ref{eq8}, the gain of the JTWPA can be calculated, as shown in Figure \ref{fig4}(b). Phase matching, $\kappa=0$ is achieved at two frequencies, $\omega_s/2\pi\approx3.1$ and $\approx 8.1$ GHz, as indicated by the vertical dashed lines. The corresponding gain, shown in Figure \ref{fig4}(b), reaches its maximum at these frequencies due to the exponential dependence of gain on $l$. For $\omega_s$ close to $\omega_p$ where $\kappa\neq 0$, the gain is limited due to the quadratic dependence of the gain on $l$.

\begin{figure}
\includegraphics[scale=0.43]{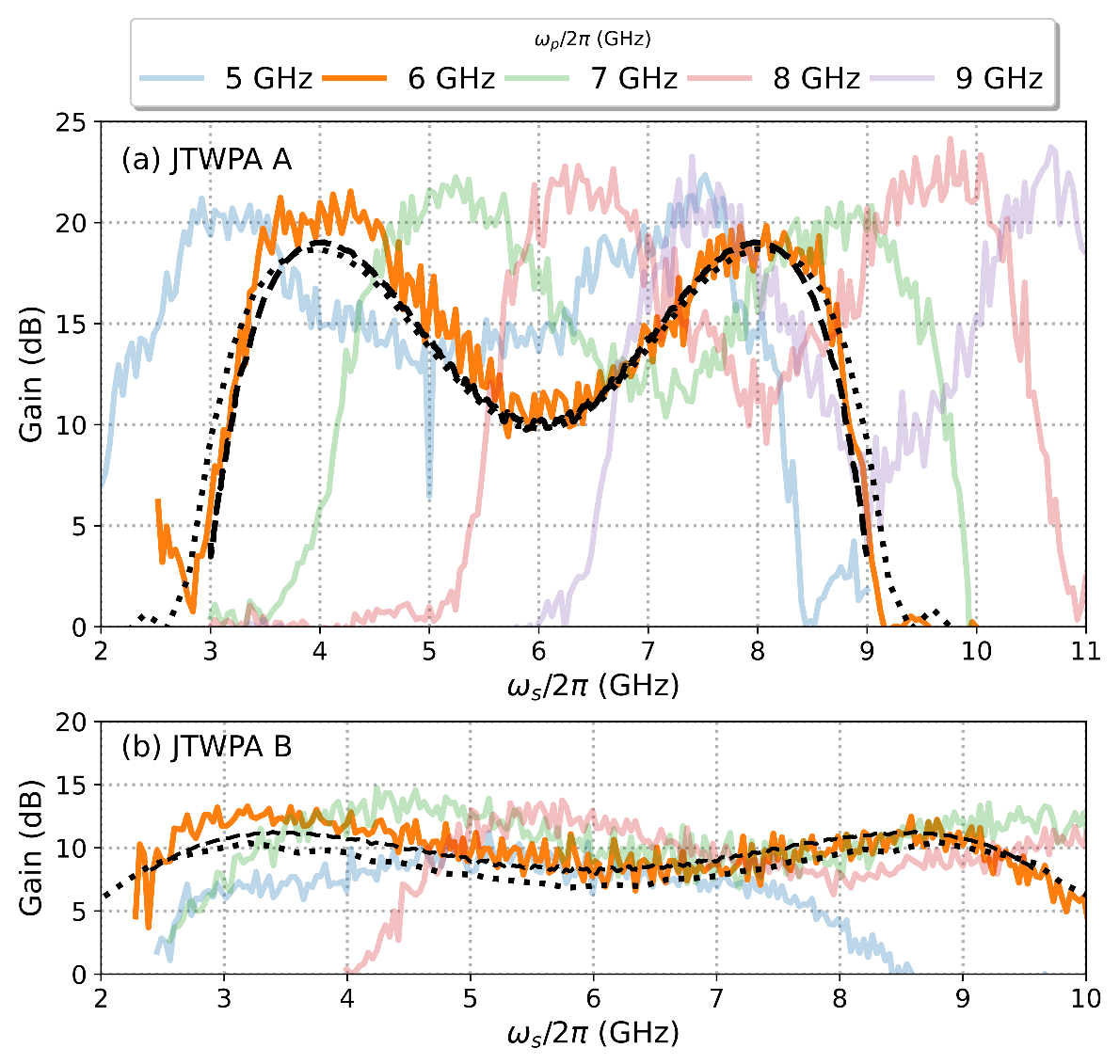}
\caption{\label{fig5}Measured gain for JTWPA A (a) and B (b) for different $\omega_p$. Gain measurements in (a) for JTWPA A show $\sim$ 20 dB of gain over an instantaneous bandwidth of 3 GHz for $\omega_p/2\pi = 6$ GHz and an \textit{in situ} tunable bandwidth of 8 GHz. Dashed lines are calculations of the gain from Eq. \ref{eq6}. Dotted lines are WRSpice simulations of the gain.}
\end{figure}

The key results of this work are the measured signal gains of the JTWPAs across a broad range of $\omega_s$, presented in Figure \ref{fig5}(a) JTWPA A and (b) JTWPA B, with $\omega_p/2\pi=$ 5 (blue line), 6 (orange line), 7 (green line), 8 (red line), and 9 GHz (purple line). For these measurements, $\Phi/\Phi_0=0.475$ for JTWPA A and 0.45 for JTWPA B. $P_p$ was adjusted to the maximum before $\theta_{\rm NL}$ deviated from linearity, as shown in Figure \ref{fig3}(c). For representative measurements of both JTWPAs (solid orange lines), the pump conditions were $\omega_p/2\pi = 6$ GHz and  $P_p$ = -78±1.5 dBm (JTWPA A) and -75±1.5 dBm (JTWPA B).

A unique aspect of this JTWPA design is its tunability in $\omega_p$, which is not constrained by dispersion engineering resonant circuits or photonic bandgap features, set during fabrication. The absence of such dispersive features within the JTWPA bandwidth reduces potential gain ripples caused by impedance mismatches associated with the introduction of these circuits. These results demonstrate the JTWPA's flexibility in tuning the center frequency of the gain bandwidth across a broad range, achieving an ultimate tunable bandwidth of 8 GHz. Moreover, the data show that maximum phase-matched gain occurs at $\omega_s$ well separated from $\omega_p$, enabling effective pump isolation using filters or diplexers instead of bulky isolators \cite{twpaperiphckts1,Rochtwpaiso}.

The dashed lines in Figure \ref{fig5} represent gain calculations based on Eq. \ref{eq6}, derived from measurements of the chromatic dispersion and nonlinear propagation constant for the representative conditions of JTWPA A and B. The measured gains are in good agreement with the theoretical predictions, with no free parameters used. Numerical time-domain simulations utilizing the circuit simulator WRspice \cite{wrspce_pub,wrspice}, were used to simulate the JTWPA circuit in Figure \ref{wrckt} with the circuit parameters in Table \ref{table1}. The simulated gain for both JTWPAs shown in Figure \ref{fig5} (dotted lines) are in good agreement with what was observed experimentally.

\begin{figure}
\includegraphics[scale=0.58]{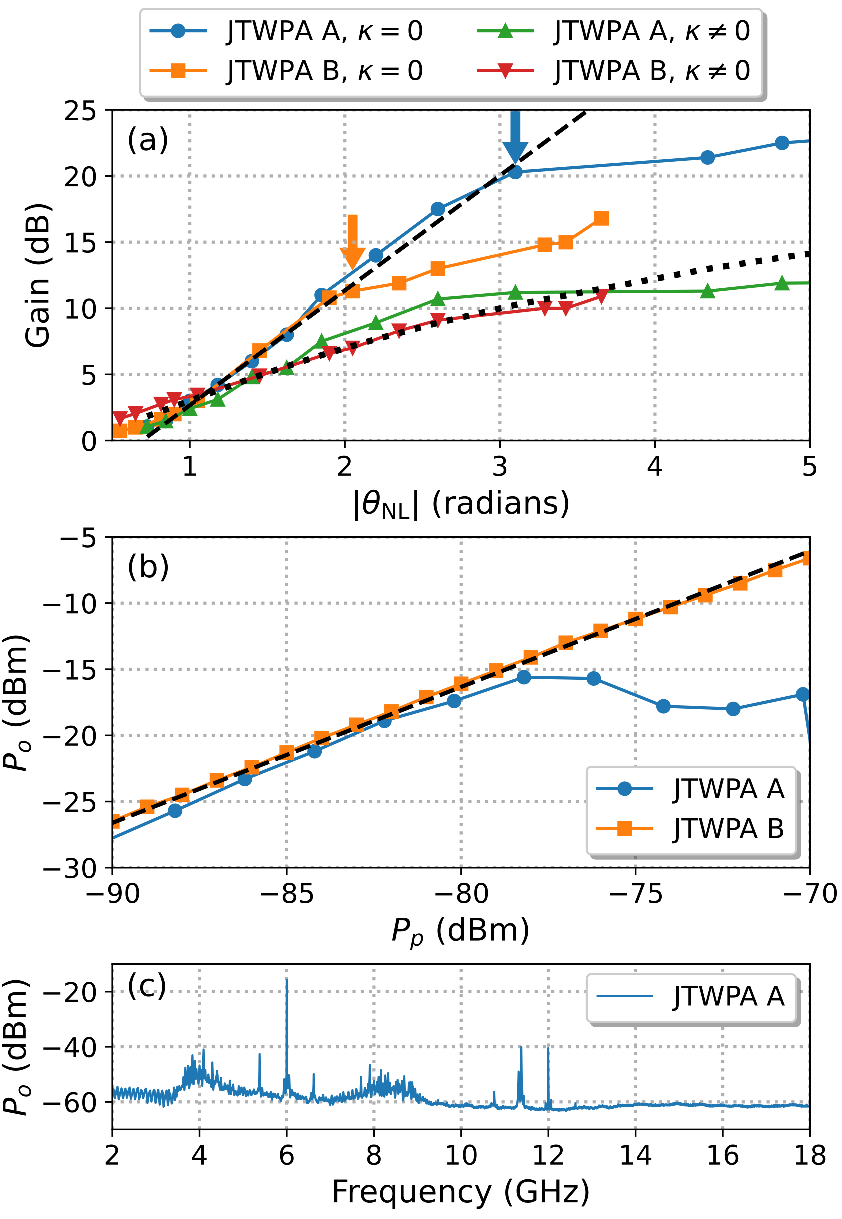}
\caption{\label{fig6}(a) Measured gain versus $\theta_{\rm NL}$ of the pump for JTWPA A (blue circles) and JTWPA B (orange squares) at phase matching. Measured gain near the pump, where the JTWPA is not phase-matched, JTWPA A (green triangles) and JTWPA B (red inverted triangles). Dashed line illustrates expected phase matched exponential dependence of the gain versus $\theta_{\rm NL} = \alpha_p l$. Dotted line illustrates a quadratic gain dependence expected for mismatched phase. Arrows show bias points $P_p$ = $-78$ dBm (blue arrow) for JTWPA A and $P_p$ = $-75$ dBm (orange arrow) for JTWPA B. (b) Spectrum analyzer measured $P_o$ versus $P_p$ which shows a saturation of the pump for higher $P_p$ in both JTWPAs. Dashed line illustrates proportionality between $P_o$ and $P_p$ (c) Spectrum analyzer trace of $P_o$ taken at $P_p$ = $-76$ dBm for JTWPA A showing higher order parametric mixing products in frequency which may be depleting the pump.}
\end{figure}

The pump conditions for both JTWPAs in Figure \ref{fig5} were carefully selected to ensure that $\theta_{\rm NL}$ for a given $\omega_p$ maintained a linear relationship with $P_p$, as described by Eq. \ref{eq8}. Although higher $P_p$ resulted in a greater experimentally observed gain, it did not follow an exponential dependence on $P_p$  or the associated $\theta_{\rm NL}$. Figure \ref{fig6}(a) shows the measured signal gain in the phase-matching region when $\kappa \approx 0$ at $\omega_s/2\pi = 4$ GHz versus $\theta_{\rm NL}$ for JTWPA A (blue circles) and JTWPA B (orange squares) with $\omega_p/2\pi = 6$ GHz. The dashed line represents the expected exponential gain dependence on $\theta_{\rm NL}$ from Eq. \ref{eq6}. The measured signal gains near $\omega_p$, where the JTWPA is not phase-matched $\kappa \neq 0$, are shown for JTWPA A (green triangles) and JTWPA B (red inverted triangles). In this case, the signal gain matches a quadratic dependence on $\theta_{\rm NL}$, as illustrated by the dotted line from Eq. \ref{eq6}. For both JTWPAs, under phase-matching conditions at low $P_p$ the gain increases with a quadratic like dependence. For larger $\theta_{\rm NL}$, the signal gain follows an exponential trend up to $\theta_{\rm NL}=3.1$ radians ($P_p$ = $-78$ dBm) for JTWPA A (blue arrow) and $\theta_{\rm NL}=2.1$ radians ($P_p$ = $-75$ dBm) for JTWPA B (orange arrow). Beyond these values, the signal gain reverts to a quadratic dependence indicating a saturation in the gain of the JTWPA versus $P_p$.

Measurements of the uncalibrated output power $P_o$ versus $P_p$ at $\omega_p/2\pi = 6$ GHz were performed using a spectrum analyzer, Figure \ref{fig6}(b) for JTWPA A (blue circles) and JTWPA B (orange squares). At low $P_p$, the increase in input $P_p$ is mirrored by a corresponding increase in $P_o$. However, for $P_p > -78$ dBm for JTWPA A and $P_p > -75$ dBm for JTWPA B, $P_o$ does not rise at the same rate as the input. This suggests that the pump is being depleted as it propagates along the length of the JTWPA. In the longer JTWPA A, $P_o$ saturates at a level 3 dB lower than in the shorter JTWPA B, despite both JTWPAs having nominally identical circuit parameters. This indicates that pump depletion is occurring along the length of the JTWPA, with the shorter JTWPA B being able to support higher $P_p$, achieving more gain per unit length. The spectrum of the output for JTWPA A at $P_p=-76$ dBm is shown in Figure \ref{fig6}(c), where several peaks are observed, including one at the second harmonic of $\omega_p$. Second harmonic generation was not expected in this JTWPA due to the absence of a second order nonlinearity. However, second harmonic generation in JTWPA circuits designed to suppress the effects of second order nonlinearities have been experimentally observed and linked to variations in Josephson junction critical currents \cite{twpasim4_pumpharmonics}. We believe the observed pump depletion is linked to the generation of higher order parametric mixing products, the exact method is unknown and further investigations are necessary. 

The data in Figure \ref{fig3}(c) and Figure \ref{fig6} provide clear evidence of pump depletion, as seen in the nonlinear increase of $\theta_{\rm NL}$ with $P_p$ and the breakdown of the exponential dependence of gain on $\theta_{\rm NL}$.  The phase-matched signal gain data in Figure \ref{fig6}(a) demonstrates a transition from exponential to quadratic dependence at large $\theta_{\rm NL}$ as the pump becomes depleted. For JTWPAs pumped with $P_p$ above this transition the data suggests that in a fraction of the JTWPA's length, phase matching is maintained due to a nearly constant pump amplitude, resulting in gain which has grown exponentially along $z$. As the pump propagates higher order parametric products are generated, and increase in amplitude versus $z$, and eventually deplete the pump. At this point phase matching deteriorates, and the signal gain versus $z$ reverts to a quadratic dependence as observed in experiment. These effects limit how strong the JTWPA can be pumped, and how efficient is the gain versus $z$ of the JTWPA.

\begin{figure}
\includegraphics[scale=0.58]{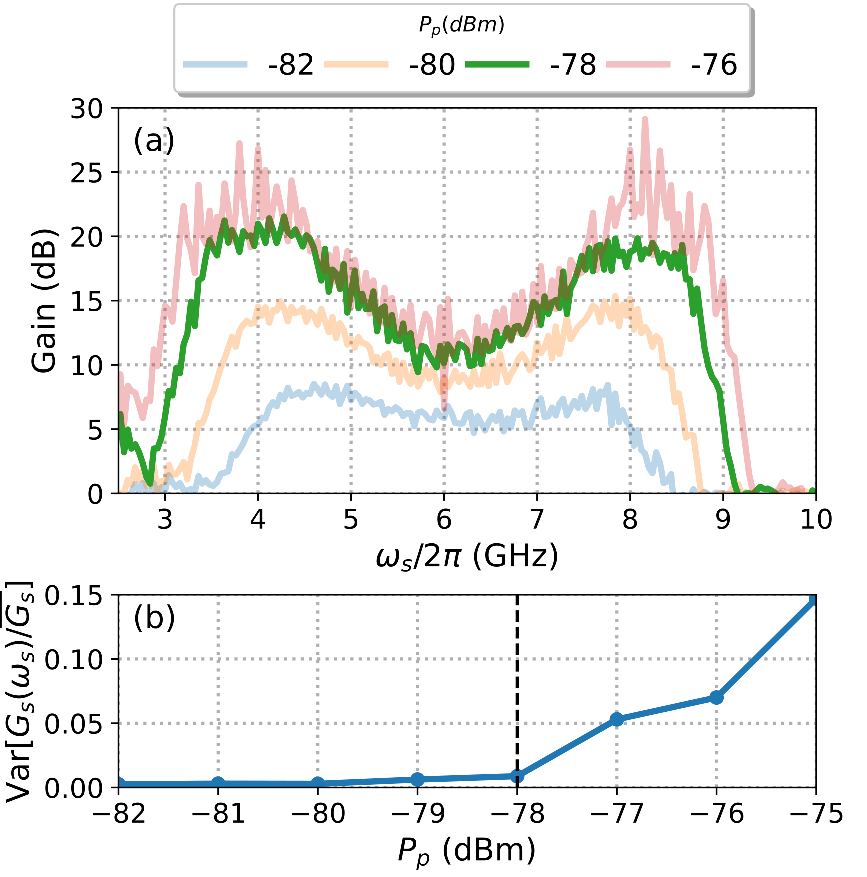}
\caption{\label{fig7}(a) JTWPA A Gain versus $\omega_s/2\pi$ for $\omega_p/2\pi = 6$ GHz and $P_p$: -82 dBm (blue line), $-80$ dBm (orange line), $-78$ dBm (green line), and $-76$ dBm (red line). (b) Plot of the variance of the normalized gain versus $P_p$. For $P_p > -78$ dBm (vertical dashed line) the variance in the normalized gain increases significantly and results in sizable ripples in the gain.}
\end{figure}

Key challenges with JTWPAs are limited signal gain and the presence of gain ripples within the usable bandwidth. It is crucial to ensure that the JTWPA is not pumped with excessively large $P_p$, which could lead to only a portion of the JTWPA being phase-matched, thereby limiting the overall gain. Since the propagation constant and phase matching of the JTWPA are highly dependent on $P_p$, gain ripples can form in the pass-band due to non-uniform pump amplitude along $l$. Figure \ref{fig7}(a) shows the signal gain of JTWPA A versus $\omega_s$ at various $P_p$: -82 dBm (blue), $-80$ dBm (orange), $-78$ dBm (green), and $-76$ dBm (red) for $\Phi/\Phi_0 = 0.475$. As seen in the figure, the gain ripples increase with $P_p$. Since the overall gain is strongly dependent on $P_p$, the variance of the gain $G_s(\omega_s)$ normalized to the mean gain in frequency in the phase-matching region Var[$G_s(\omega_s)/\overline{G_s}]$ of the JTWPA was calculated for each $P_p$ and plotted in Figure \ref{fig7}(b). The dashed line marks $P_p = -78$ dBm where pump saturation occurs, at which point the JTWPA is no longer fully phase-matched, and this is where a notable increase in gain ripples with increasing $P_p$ is observed.

\begin{figure}
\includegraphics[scale=0.58]{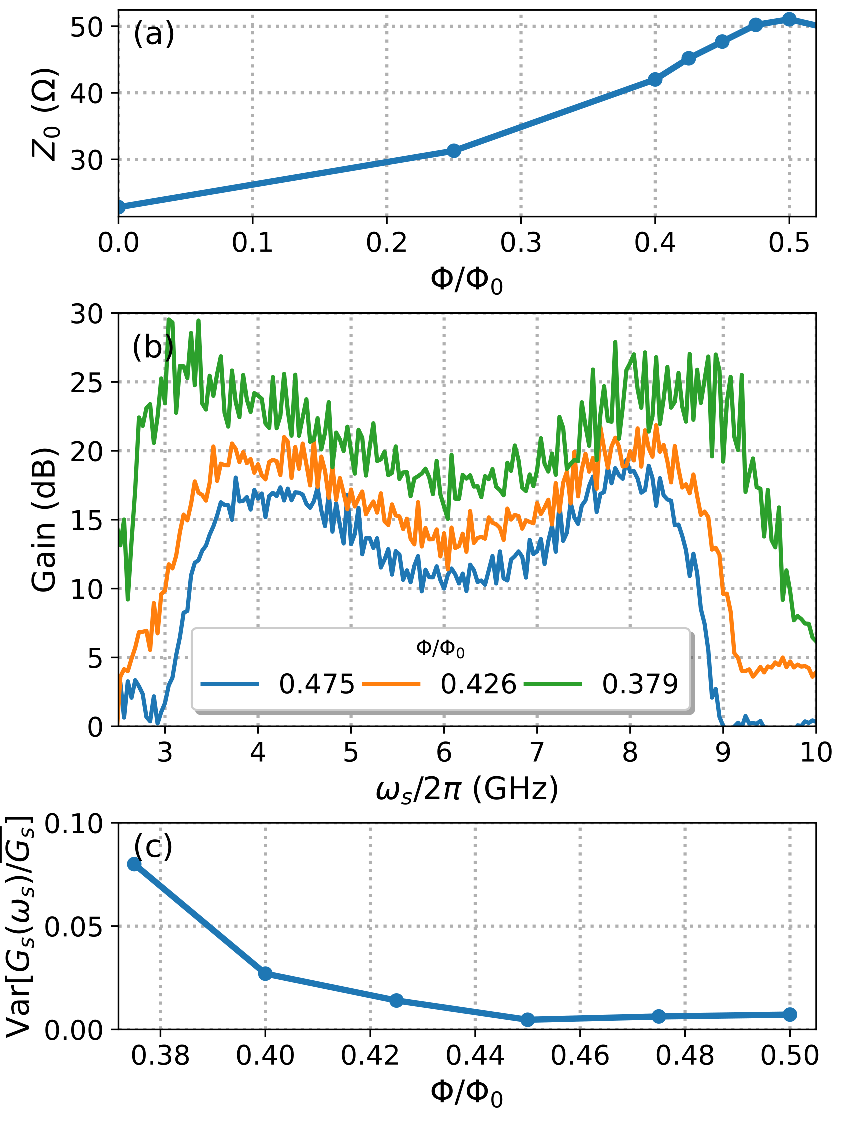}
\caption{\label{fig8}(a) Characteristic impedance of JTWPA A versus tuning $\Phi/\Phi_0$. (b) Measurements of the gain at $\omega_p/2\pi = 6$ GHz for $\Phi/\Phi_0 = 0.475$ (blue line), 0.426 (orange line), and 0.379 (green line). $P_p$ was adjusted to achieve an equal mean gain in the phase-matching region. Traces are offset vertically by 4 dB clarity. (c) Variance in the normalized gain versus $\Phi/\Phi_0$. As $Z_0$ is tuned away from 50 $\Omega$ the gain variance and ripple increase substantially.}
\end{figure}

An impedance mismatch between the JTWPA and the 50 $\Omega$ environment can create standing waves along the JTWPA transmission line, leading to ripples in the gain. The characteristic impedance $Z_0$ of JTWPA A, versus $\Phi/\Phi_0$, was determined from transmission measurements of the dispersion in Figure \ref{fig3}(a) and from test resonator measurements to determine $C_{\rm gnd}$ as shown in Figure \ref{fig8}(a). At $\Phi/\Phi_0 = 0.475$ JTWPA A is matched to the readout with 50 $\Omega$ impedance. When the JTWPA A is tuned to $\Phi/\Phi_0 = 0$, $Z_0 = $ 22 $\Omega$ and results in an impedance mismatch of the JTWPA at both ends. This impedance mismatch results in multiple reflections and a standing wave in the transmission with a 2.5 dB ripple, which was experimentally observed in Figure \ref{fig2}(b),(c). 

Figure \ref{fig8}(b) shows measurements of the gain of JTWPA A at $\omega_p/2\pi = 6$ GHz for $\Phi/\Phi_0 = 0.475$ (blue line), 0.426 (orange line), and 0.379 (green line). In these measurements, $P_p$ was adjusted to achieve an equal mean gain in the phase-matching region, with the traces in Figure \ref{fig8}(b) offset vertically by 4 dB for clarity. The variance of the normalized gain Var[$G_s(\omega_s)/\overline{G_s}]$ in the phase matched region is plotted in Figure \ref{fig8}(c) versus $\Phi/\Phi_0$. The results show that when the JTWPA's $Z_0$ is tuned away from $50 \Omega$, the gain ripple increases as expected.

\begin{figure}
\includegraphics[scale=0.58]{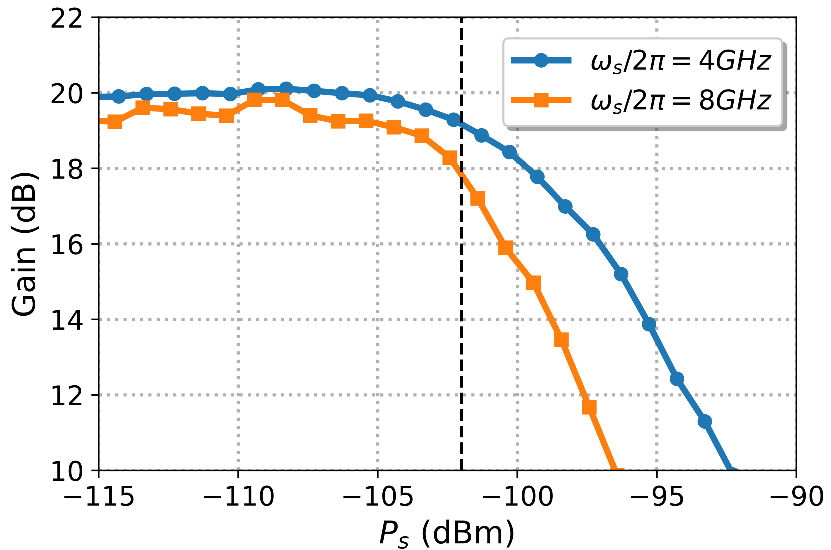}
\caption{\label{fig9}Measurements of the saturation power in the phase matching region at $\omega_s/2\pi = $ 4 (blue circles) and 8 GHz (orange squares) indicated by the 1 dB compression point (vertical dashed line at $-102$ dBm) for JTWPA A pumped at $\omega_p/2\pi = 6$ GHz, $Pp = -78$ dBm, and $\Phi/\Phi_0 = 0.475$.}
\end{figure}

The saturation power of JTWPA A was determined by measuring the signal gains versus signal power $P_s$ at $\omega_s/2\pi=$ 4 (blue circles) and 8 GHz (orange squares) with $\omega_p/2\pi = $ 6 GHz and $P_p = -78$ dBm shown in Figure \ref{fig9}. For phase-matched signal gains of approximately 20 dB, a 1 dB compression in gain, indicating the saturation power, was observed at input signal powers of $-102$ dBm (vertical dashed line). WRspice simulations of the 1 dB compression point confirm that the output signal power can be within 4 dB of $P_p$ before significant gain compression occurs. Measurements of JTWPA A with larger $P_p$ had saturation powers, up to $-98$ dBm. JTWPAs pumped with larger $P_p$  resulted in inefficient phase matching, increased gain ripples, and additional noise from the JTWPA.

\section{\label{noise}Noise Measurements}

\begin{figure}
\includegraphics[scale=0.6]{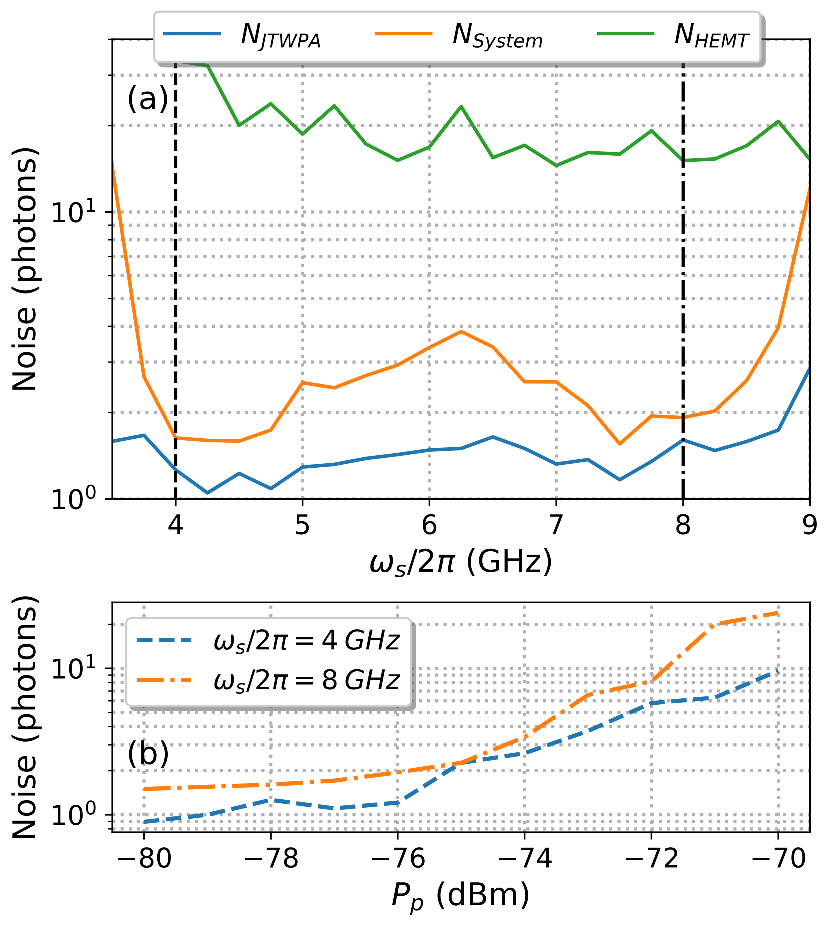}
\caption{\label{fig10} Noise measurements: (a) Effective HEMT noise (green line), system noise (orange line), and JTWPA noise (blue line) in photon units utilizing the signal-to-noise ratio improvement method for JTWPA A. (b) Measured JTWPA noise in the phase matching region at $\omega_s/2\pi = 4$ (blue dash line) and 8 GHz (orange dash dotted line) versus $P_p$. Noise increases above $Pp = -76$ dBm as the number of parametric conversion processes increase.}
\end{figure}

Noise measurements of the JTWPA were conducted using the signal-to-noise ratio improvement method \cite{snrimprov}. The initial step in characterizing the JTWPA's noise involved characterizing the effective noise of the High-electron mobility transistor (HEMT) amplifier readout chain. A broadband thermal noise source whose temperature was varied between 150 mK to 4K was utilized to perform a Y-factor measurement where the power spectral density was measured with a spectrum analyzer \cite{pozar}. The measured noise power was then analyzed and fitted to the noise model

\begin{eqnarray}\label{eq9}
P_{noise}(\omega, T) = \nonumber\\
\left[\frac{\hbar\omega}{2}\left(\text{coth}\frac{\hbar\omega}{2k_BT}\right)+k_B T_{\rm HEMT}(\omega)\right]G(\omega)B,
\end{eqnarray}

\noindent where $T_{\rm HEMT} (\omega)$ is the effective noise temperature of the HEMT, $G(\omega)$ is the system gain, and $B$ is the measurement bandwidth. The first term in Eq. \ref{eq9} is the noise emitted from the noise source at frequency $\omega$ when heated to a temperature $T$. The measured noise in photon units $N_{\rm HEMT}=k_BT_{\rm T}/\hbar\omega_s$ is shown in Figure \ref{fig10}(a) (green line). This effective noise originates from the HEMT amplifier and additional losses before the HEMT from cables and isolators used in the measurement setup. 

After characterizing the noise of the output line, the JTWPA is switched into the measurement chain with a cryogenic RF switch, and both the signal and noise gain are measured with the JTWPA switched on and off. The transmission when the JTWPA is on, referenced to the thruline measurement, provides the overall JTWPA gain $G_{\rm JTWPA}$, which includes JTWPA losses. The increase in the measured noise power when the JTWPA is switched on compared to when it is switched off provides the noise gain $G_{\rm noise}$. The system and JTWPA noise temperatures can be calculated: $T_{\rm system}=G_{\rm noise} T_{\rm HEMT}⁄G_{\rm JTWPA}$  and $T_{\rm JTWPA}=T_{\rm HEMT} (G_{\rm noise}-1)⁄G_{\rm JTWPA}$. Figure \ref{fig10}(a) shows the noise of JTWPA A (blue line) and system noise (orange line) in photon units, for $\Phi/\Phi_0 = 0.475$, $\omega_p/2\pi = 6$ GHz, and $P_p = -78$ dBm. In the phase-matched regions around $\omega_s/2\pi = 4$ and 8 GHz, JTWPA A demonstrates near quantum-limited noise performance, with noise $\sim 1.5$ photons. Due to the high gain of JTWPA A, the system noise is below 2 photons in the phase-matching regions.

Measurements of the noise of JTWPA A in regions of phase matching at $\omega_s/2\pi = 4$ and 8 GHz as $P_p$ is varied is shown in Figure \ref{fig10}(b). For low powers, the added noise of the JTWPA is near quantum limited around the lower lobe in the gain centered at $\omega_s/2\pi = 4$ GHz (See Figure \ref{fig5}(a)), and for gain centered at $\omega_s/2\pi = 8$ GHz the noise is less than 2 photons. The low added noise by the JTWPA is attributed to the low loss and efficient phase matching of the JTWPA. The mismatch in noise between the two phase-matched gain lobes is due to asymmetry in the gain and losses at those frequencies in the JTWPA. The noise is shown to increase with $P_p$ as the number of parametric conversion products increases \cite{floquet} Figure \ref{fig6}(c). The increase in noise with $P_p$ highlights the importance of not over pumping the JTWPA, in this case to $P_p \gtrsim -78$ dBm.

\section{\label{discussion}Discussion}

We have presented a JTWPA that employs a novel inverse Kerr phase-matching technique. The JTWPA circuit incorporates an asymmetric SQUID unit-cell, enabling control over the amplitude and sign of the Kerr nonlinearity. Combined with the transmission line's chromatic dispersion, this design achieves phase matching for a four-wave mixing parametric processes while suppressing second-order processes. A key advantage of the JTWPA design presented is the ability to tune the Kerr nonlinearity, the characteristic impedance of the transmission line, and the pump frequency using an external magnetic field. This tunability allowed for the identification of an optimal operating point, which maximizes gain and ensures efficient exponential growth with amplifier length and pump power, while minimizing gain ripple and added noise. When optimally tuned, the JTWPA achieved 20 dB of gain over a 3 GHz instantaneous bandwidth, with a tunable pump bandwidth of 8 GHz, and minimal gain ripple. The noise performance was near the quantum limit, largely due to the use of low-loss dielectrics in the design.
Given the increasing interest in JTWPA technology, JTWPA transmission line circuits with \textit{in situ} control over nonlinearities could reveal novel possibilities as quantum devices which can produce or manipulate non-classical states of light or more versatile parametric amplifiers with characteristics such as reverse isolation which aids in scalability of readout for quantum computing systems.

\begin{acknowledgments}
Thank you to M. Gershenson, Ted White, and Joe Aumentado for useful discussions. I thank S. Tolpygo and W. Oliver for first generation prototypes of the JTWPA which were fabricated at Lincoln Laboratories. This work was supported in part by NSF grants DMR-1838979, OSI-2328774 and TI-2025848. Device characterization was performed in the University of Massachusetts Boston Quantum Core facility which is funded through a grant from the Massachusetts Technology Collaborative matching grant program. M.T. Bell acknowledges funding from Google, through a Google Academic Research Award. Thank you to F.Vischi for critical reading of the manuscript.
\end{acknowledgments}

\appendix
\counterwithin{figure}{section}

\section{Measurement Setup}

\begin{figure}
\includegraphics[scale=1.25]{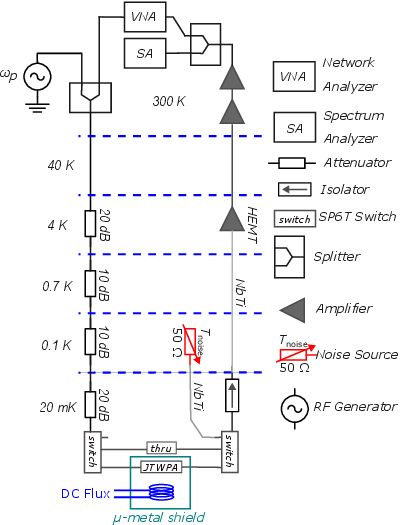}
\caption{\label{msetup}Dilution refrigerator measurement setup used to characterize JTWPAs}
\end{figure}

The measurement system for the dilution refrigerator included a standard set of attenuated input coaxial cables, which carried both the pump and signal, combined at room temperature. The output readout line incorporated RF isolators, a cryogenic HEMT amplifier, and room temperature amplification. All measurements were conducted using a vector network analyzer and a spectrum analyzer. Two cryogenic RF switches were employed to alternate between the JTWPA, a thruline, and a broadband noise source. The thruline, which consisted of an on-chip version in an identical package to the JTWPA, was used to calibrate the system's transmission magnitude and phase response, ensuring that these could be excluded from the JTWPA's transmission characterization. The JTWPA, along with an external magnetic solenoid, was placed inside a cryogenic high-$\mu$ magnetic shield. The external magnetic solenoid allowed for tuning of the JTWPA’s operating point.

 For noise characterization of the JTWPA, a broadband thermal noise source was used, consisting of a NiCr 50 $\Omega$ load, a calibrated thermometer, and a heater, all mounted on a copper bracket. This setup was attached to the intermediate cold plate of a dilution refrigerator. The thermal noise source had a weak thermal link to the cold plate, allowing for continuous tunability of its temperature from 150 mK to 4 K without affecting the temperature of the mixing chamber. A NbTi superconducting coaxial cable connected the noise source to the RF switch on the output line, providing thermal isolation and minimizing signal loss. This configuration enabled calibration of the effective noise temperature and gain of the output line, referenced to the cryogenic RF switch port.

 \section{WRspice Simulations}

\begin{figure}
\includegraphics[scale=0.95]{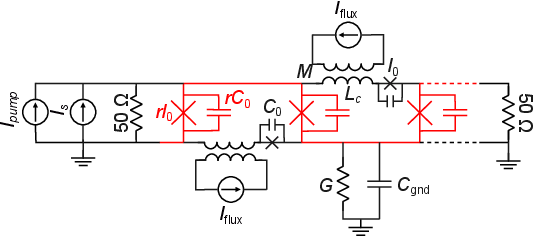}
\caption{\label{wrckt}JTWPA circuit used in WRspice simulations. Magnetic flux tuning of the JTWPA was introduced by a tuning current $I_{\rm flux}$ coupled to the asymmetric SQUIDs through a mutual inductance $M$.}
\end{figure}

WRspice circuit simulations \cite{wrspice,wrspce_pub} were used to model the gain of the JTWPA, with the circuit schematic shown in Figure \ref{wrckt}. The simulations utilized JTWPA circuit model parameters from Table \ref{table1}. Dielectric loss was represented as conductance $G$ in the simulations. To introduce a tuning $\Phi$, a geometric loop inductance $L_c$ in the asymmetric SQUID loops were coupled via a mutual inductance $M$ to a flux bias line with a tuning current $I_{\rm flux}$. Measurements of the periodicity of $S_{21}$ phase with respect to $I_{\rm flux}$ were used to determine $I_{\rm flux}$ at full frustration.

A simulation of $\angle S_{21}$ in WRspice versus pump current $I_{\rm pump}$ was used to map the experimentally measured $\angle S_{21}$ at $\omega_p/2\pi = 6$ GHz, $P_p = -78$ dBm for JTWPA A to the $I_{\rm pump}$ values in WRspice. The WRspice simulations of the gain bandwidth for both JTWPAs, shown as dotted lines in Figure \ref{fig5}, exhibit good agreement with the experimental results.

\clearpage

\bibliography{twpapaper}

\end{document}